\documentclass[english]{article}
\usepackage[T1]{fontenc}
\usepackage[latin9]{inputenc}
\usepackage{amssymb}
\usepackage{esint}
\usepackage{babel}
\begin{document}

\title{\textbf{A Machian Request for the Equivalence Principle in Extended
Gravity and non-geodesic motion}}

\author{\textbf{$^{1}$Ignazio Licata, $^{2}$Christian Corda, $^{3}$Elmo
Benedetto}}

\maketitle
\begin{center}
$^{1}$Institute for Scientific Methodology, Palermo, Italy; School
of Advanced International Studies on Theoretical and non Linear Methodologies
of Physics, Bari, I-70124, Italy; $\emph{e-mail}$: $ignaziolicata@ejtp.info$;
\par\end{center}

$^{2}$Dipartimento di Fisica, Scuola Superiore di Studi, Universitari
e Ricerca \textquoteleft{}Santa Rita\textquoteright{}, Via Tagliamento
45, 00188 Roma, Italy; Austro-Ukrainian Institute for Science and
Technology, Institut für Theoretishe Physik, Technische Universität,
Wiedner Hauptstrasse 8-10/136, A-1040, Wien, Austria; International
Institute for Applicable Mathematics \& Information Sciences (IIAMIS),
B.M. Birla Science Centre, Adarsh Nagar, Hyderabad- 500 463, India;
$\emph{e-mail}$: $cordac.galilei@gmail.com$;

$^{3}$Department of Engineering, University of Sannio, Piazza Roma
21, 82100 Benevento, Italy; $\emph{e-mail}$: $elmobenedetto@libero.it$;
\emph{orcid: 0000-0002-5920-0221}; 
\begin{abstract}
Starting from the origin of Einstein general relativity (GR) the request
of Mach on the theory's structure has been the core of the foundational
debate. That problem is strictly connected with the nature of the
mass-energy equivalence. It is well known that this is exactly the
key point that Einstein used to realize a metric theory of gravitation
having an unequalled beauty and elegance. On the other hand, the current
requirements of particle physics and the open questions within extended
gravity theories request a better understanding of Equivalence Principle
(EP). The MOND theory by Milgrom proposes a modification of Newtonian
dynamics and we consider a direct coupling between the Ricci curvature
scalar and the matter Lagrangian showing that a non geodesic ratio
$m_{i}/m_{g}$ can be fixed and that Milgrom acceleration is retrieved
at low energies.
\end{abstract}

\section{Introduction}

It is well known that \emph{The Science of Mechanics} by Ernst Mach
had a strong influence on Einstein and played an important role in
the development of GR \cite{key-1}. In Newton's \emph{Philosophiae
Naturalis Principia Mathematica}, accelleration is considered as absolute.
In the famous Gedankenexperiment of the rotating bucket filled with
water, Newton deduced the existence of an absolute rotation by observing
the curved surfaces on the water.

The aim of Newton was to explain the inertia through a sort of resistance
to motion in the absolute space which, in this way, comes to be an
agent and not a mere physical theater of coordinates, although unspecified.
The first thinker to question the Newtonian reasoning was the philosopher
George Berkeley in his \emph{De Motu}, published in 1721 and he can
be considered the precursor of Mach. Indeed, after more than 150 years,
Mach proposed a radical criticism of Newton's absolute space and he
concluded that the inertia would be an interaction that requires other
bodies to manifest itself, so that it would make no sense in a Universe
consisting of just a single mass. According to Mach, there is a total
relational symmetry and every motion, uniform or accelerated, makes
sense only in reference to other bodies. Therefore, following the
so called \emph{Mach Principle}, the inertia of a body is not an intrinsic
property and depends on the mass distribution in the rest of the Universe.
Einstein was very fascinated by Mach reasoning but it is widely acknowledged
that Mach Principle is not fully incorporated into relativistic field
equations \cite{key-2}. The challenge of a \emph{Machian physics}
was accepted several times (though less than expected) in the context
of both classical and quantum. Here we recall the Narlikar's theory
with variable mass derived from Wheeler-Feynman-like action at a distance
theory \cite{key-3,key-4}. Sciama's theory requires to get the inertia
as ``gravitational closeness'' (and the perfect equivalence) under
the precise cosmological condition $G\rho\frac{r^{2}}{c^{2}}=1$ where
$r$ is the radius of the universe, $\rho$ the density, $c$ is the
speed of light and $G$ the gravitational constant \cite{key-5}.
In quantum contexts, and in Higgs times, the problem becomes more
complex \cite{key-6,key-7,key-8,key-9,key-10}.

\section{The Physics Under the Metric}

Einstein has often stated that some Machian effects are present in
GR. In particular, in the famous Lectures of 1921 \cite{key-11} he
states that it showed the following effects:

1) \emph{The inertia of a body must increase when ponderable masses
are piled up in its neighbourhood}.

2) \emph{A body must experience an accelerating force when neighboring
masses are accelerated and the force must be in the same direction
as that acceleration}.

3) \emph{A rotating hollow body must generate inside of itself a Coriolis
field which deflects moving bodies in the sense of the rotation and
a radial centrifugal field as well}.

Let us follow the Einstein reasoning. By considering the geodesic
equation

\begin{equation}
\frac{d^{2}x_{\mu}}{ds^{2}}+\Gamma_{\alpha\beta}^{\mu}\frac{dx_{\alpha}}{ds}\frac{dx_{\beta}}{ds}=0
\end{equation}
Einstein worked in the weak- field approximation and the metric he
found to represent the gravitational field due to a distribution of
small masses corresponding to a density $\sigma$ and having small
velocities $\frac{dx_{i}}{ds}$ can be written as

\begin{equation}
\begin{array}{c}
g_{00}=1-\frac{2G}{c^{2}}\int\frac{\sigma dV}{r}\\
\\
g_{0i}=\frac{4G}{c^{2}}\int\frac{dx_{i}}{ds}\frac{\sigma dV}{r}\\
\\
g_{ij}=-\delta_{ij}\left(1+\frac{2G}{c^{2}}\int\frac{\sigma dV}{r}\right)
\end{array}
\end{equation}
The equation of motion in this field becomes

\begin{equation}
\frac{d}{dx^{0}}\left[\left(1+\bar{\sigma}\right)v\right]=\nabla\bar{\sigma}+\frac{\partial A}{\partial x^{0}}+\left(\nabla\wedge A\right)\wedge v
\end{equation}
where 
\begin{equation}
\begin{array}{c}
\bar{\sigma}\equiv\frac{G}{c^{2}}\int\frac{\sigma dV}{r}\\
\\
A=\frac{4G}{c^{2}}\int\frac{\sigma vdV}{r}.
\end{array}
\end{equation}
Einstein interpreted it by saying that the inertial mass is proportional
to $1+\bar{\sigma}$ and therefore increases when ponderable masses
approach the test body

\begin{equation}
m_{i}=m_{g}\left(1+\frac{G}{c^{2}}\int\frac{\bar{\sigma}dV}{r}\right)
\end{equation}
Many physicists believe, according to C. Brans \cite{key-12}, that
only the second and third effect are contained in GR. At first glance
it seems that, if the Einstein interpretation is right, the EP is
violated but we emphasize that all bodies with different inertial
masses are still falling with the same acceleration in a gravitational
field. In \cite{key-13} the author analyzes what he calls \emph{Modified
Mach Principle} in the context of an expanding universe. He suggests
the following definitions for the inertial mass within and beyond
the bulge of galaxies as

\begin{equation}
\begin{array}{ccc}
m_{i}=C &  & r\leq R_{0}\\
\\
m_{i}=\frac{C^{\prime}}{r}=m_{g}\frac{R_{0}}{r} &  & r>R_{0}
\end{array}
\end{equation}
where $C$ and $C^{\prime}$ are constants and he calls the first
one as inertial mass versus gravitational interaction within the bulge,
and the second one as inertial mass versus cosmological expansion
beyond the bulge. It would seem that the introduction of a genuine
Mach's principle implies a re-introduction of the distinction between
inertial mass and gravitational mass, hidden under the metric of GR
and the strong form of the EP. Let recall that the equivalence between
$m_{i}$ and $m_{g}$ is the axiomatic and constructive keystone of
GR. This raises the problem of the interpretation of the formalism
able to establish the EP on the physical meaning of the relationship
between $m_{i}$ and $m_{g}$.

\section{Inertial and Gravitational Mass}

The nature of dark matter is one of the unsolved mysteries in cosmology
since C. Zwicky measured the velocity dispersion of the Coma cluster
of galaxies \cite{key-14}. Let us rewrite the following relation

\begin{equation}
m_{i}\frac{v^{2}}{r}=\frac{GM_{g}m_{g}}{r^{2}},
\end{equation}
where $m_{i}$ is a body that rotates around a gravitational mass
$M_{g}$ over a constant radius $r$. The relation

\begin{equation}
v=\sqrt[4]{GMa_{0}}
\end{equation}
is in perfect agreement with the experimental data with $a_{0}$ is
about $10^{-10}\frac{m}{s^{2}}$ \cite{key-15} - \cite{key-23}.
Therefore we write

\begin{equation}
v^{2}=\frac{GM_{g}}{r}\frac{m_{g}}{m_{i}}=\sqrt{GM_{g}a_{0}}.
\end{equation}
It follows that

\begin{equation}
\frac{m_{g}}{m_{i}}=\sqrt{\frac{a_{0}r^{2}}{GM_{g}}}.
\end{equation}
If we do not interpret $a_{0}$ from the kinematic point of view but
as a gravitational field, we can write

\begin{equation}
\frac{m_{g}}{m_{i}}=\sqrt{\frac{g_{0}}{g}}.
\end{equation}
According to Mach and his interpreters, the inertial mass of a body
arises as a consequence of its interactions with the Universe and
so we assume that

\begin{equation}
\frac{m_{g}}{m_{i}}=\mu(x)
\end{equation}
with $\mu=1$ for $\left\vert \frac{g}{g_{0}}\right\vert \gg1$ and
$\mu=\sqrt{
}$ for $\left\vert \frac{g}{g_{0}}\right\vert \ll1.$

A possible form of $\mu$ may be for example

\begin{equation}
\mu=\sqrt{\frac{g_{0}+g}{g}}
\end{equation}
where $g$ is the field generated by nearby masses.

It is easy to verify that when $g\gg g_{0}$, circular velocity decreases
in Keplerian way but when $g\ll g_{0}$ we obtain

\begin{equation}
v^{2}=\frac{GM_{g}}{r}\sqrt{\frac{g_{0}}{g}}=GM_{g}\sqrt{\frac{g_{0}}{GM_{g}}}=\sqrt{GM_{g}g_{0}}
\end{equation}
and finally

\begin{equation}
v=\sqrt[4]{GMg_{0}}.
\end{equation}
Obviously the value of $g_{0}$ that fits all the data of galaxies
rotation curves is about $10^{-10}m/s^{2}.$ From the mathematical
point of view the relations (8) and (15) coincide but from a physical
point of view the situation is different. At every point in the Universe,
the second Newtonian law is still valid even for small accelerations.

Deviations between inertial and gravitational mass as stressed by
eq. (10) can have an intriguing geometrical explanation in the framework
of $f(R)$ theories of gravity, assuming an explicit coupling between
an arbitrary function of the scalar curvature, $R$, and the Lagrangian
density of matter \cite{key-25}. The most simple situation is to
consider a weak modification of general relativity, which could be
consistent with solar system tests, which implies a direct coupling
between the Ricci curvature scalar and the matter Lagrangian \cite{key-26}.
Following \cite{key-25,key-26}, let us consider the action (for the
sake of simplicity we set $16\pi G=1$, $c=1$ and $\hbar=1$ hereafter)
\begin{equation}
S=\int d^{4}x\sqrt{-g}\left(R+\lambda R\mathcal{L}_{m}+\mathcal{L}_{m}\right),\label{eq: high order 1}
\end{equation}
which only includes a coupling between the Ricci scalar and the matter
Lagrangian, being $\lambda$ the coupling constant, with respect to
the well known canonical Einstein - Hilbert action of standard general
relativity \cite{key-27}

\begin{equation}
S=\int d^{4}x\sqrt{-g}.\label{eq: EH}
\end{equation}
Without loss of generality, we can also set $\lambda=1$. Thus, the
standard variation analysis in a local Lorentz frame enables to write
\cite{key-25,key-26} 
\begin{equation}
\begin{array}{c}
\delta\int d^{4}x\sqrt{-g}\left(R+R\mathcal{L}_{m}+\mathcal{L}_{m}\right)=\int d^{4}x\left[\delta\sqrt{-g}\left(R+R\mathcal{L}_{m}+\mathcal{L}_{m}\right)+\sqrt{-g}\delta\left(R+R\mathcal{L}_{m}+\mathcal{L}_{m}\right)\right]\\
\\
=\int d^{4}x\left[\sqrt{-g}\left(1+\mathcal{L}_{m}\right)R_{\mu\nu}-\frac{1}{2}g_{\mu\nu}\left(R+R\mathcal{L}_{m}+\mathcal{L}_{m}\right)\right]g^{\mu\nu}+d^{4}x\sqrt{-g}\left(1+\mathcal{L}_{m}\right)g^{\mu\nu}\delta R_{\mu\nu}.
\end{array}\label{eq: high order 3}
\end{equation}
 The relation between the connections and the Ricci tensor \cite{key-27}
gives \cite{key-25,key-26} 
\begin{equation}
g^{\mu\nu}\delta R_{\mu\nu}=g^{\mu\nu}\partial_{\alpha}\left(\delta\Gamma_{\mu\nu}^{\alpha}\right)-g^{\mu\alpha}\partial_{\alpha}\left(\delta\Gamma_{\mu\nu}^{\nu}\right)\equiv\partial_{\alpha}X^{\alpha},
\end{equation}
where 
\begin{equation}
X^{\alpha}\equiv g^{\mu\nu}\left(\delta\Gamma_{\mu\nu}^{\alpha}\right)-g^{\mu\alpha}\left(\delta\Gamma_{\mu\nu}^{\nu}\right).
\end{equation}
 Thus, the second integral in equation (18) results to be \cite{key-25,key-26}

\begin{equation}
\begin{array}{c}
\int d^{4}x\sqrt{-g}(1+\mathcal{L}_{m})g^{\mu\nu}\delta R_{\mu\nu}=\int d^{4}x\sqrt{-g}(1+\mathcal{L}_{m})\partial_{\alpha}X^{\alpha}=\\
\\
=\int d^{4}x\partial_{\alpha}[\sqrt{-g}(1+\mathcal{L}_{m})X^{\alpha}]-\int d^{4}x\partial_{\alpha}[\sqrt{-g}(1+\mathcal{L}_{m})]X^{\alpha}.
\end{array}
\end{equation}
A standard assumption is that fields are equal to zero at infinity.
Then we obtain \cite{key-25,key-26}

\begin{equation}
d^{4}x\sqrt{-g}(1+\mathcal{L}_{m})g^{\mu\nu}\delta R_{\mu\nu}=-\int d^{4}x\partial_{\alpha}[\sqrt{-g}(1+\mathcal{L}_{m})]X^{\alpha}.\label{calcolo 2}
\end{equation}
Let us calculate the quantity $X^{\alpha}.$ In a local Lorentz frame
it is \cite{key-25,key-26}

\begin{equation}
\bigtriangledown_{\beta}g_{\mu\nu}=\partial_{\beta}g_{\mu\nu}=0.\label{eq: delta-delta}
\end{equation}
Thus, the well known definitions of the Christofell connections \cite{key-27}
gives \cite{key-25,key-26}

\begin{equation}
\begin{array}{c}
\delta\Gamma_{\mu\nu}^{\alpha}=\delta[\frac{1}{2}g^{\beta\alpha}(\partial_{\mu}
\\
\\
=\frac{1}{2}g^{\beta\alpha}\left(\partial_{\mu}\delta_{\beta\nu}+\partial_{\nu}\delta_{\mu\beta}-\partial_{\beta}\delta_{\mu\nu}\right).
\end{array}\label{eq: g}
\end{equation}
In analogous way one gets \cite{key-25,key-26} 
\begin{equation}
\delta\Gamma_{\mu\nu}^{\nu}=\frac{1}{2}g^{\nu\beta}\partial_{\mu}(\delta g_{\nu\beta}).\label{eq: g2}
\end{equation}
Using eqs. (24) and (25) we find \cite{key-25,key-26}

\begin{equation}
g^{\mu\nu}(\delta\Gamma_{\mu\nu}^{\alpha})=\frac{1}{2}\partial^{\alpha}(g_{\mu\nu}\delta g^{\mu\nu})-\partial^{\mu}(g_{\beta\mu}\delta g^{\nu\beta})\label{eq: calcola}
\end{equation}
and 
\begin{equation}
g^{\mu\alpha}(\delta\Gamma_{\mu\nu}^{\nu})=-\frac{1}{2}\partial^{\alpha}(g_{
}\delta g^{\nu\beta}).
\end{equation}
Now, substituting in (20), we obtain \cite{key-25,key-26} 
\begin{equation}
X^{\alpha}=\partial^{\alpha}(g_{\mu\nu}\delta g^{\mu\nu})-\partial^{\mu}(g_{\mu\nu}\delta g^{\alpha\nu}).\label{eq: X2}
\end{equation}
In this way, equation ($\ref{calcolo 2}$) becomes \cite{key-25,key-26}
\begin{equation}
\begin{array}{c}
\int d^{4}x\sqrt{-g}(1+\mathcal{L}_{m})g^{\mu\nu}\delta R_{\mu\nu}=\\
\\
=\int d^{4}x\partial_{\alpha}[\sqrt{-g}(1+\mathcal{L}_{m})][\partial^{
}(g_{\mu\nu}\delta g^{\alpha\nu})-\partial^{\alpha}(g_{
}\delta g^{\nu\beta}],
\end{array}
\end{equation}
which also gives \cite{key-25,key-26} 
\begin{equation}
\begin{array}{c}
\int d^{4}x\sqrt{-g}(1+\mathcal{L}_{m})g^{\mu\nu}\delta R_{\mu\nu}=\\
\\
=\int d^{4}x\left\{ g_{\mu\nu}\partial^{\alpha}\partial_{\alpha}[\sqrt{-g}(1+
\right\} -\int d^{4}x\left\{ g_{\mu\nu}\partial^{\mu}\partial_{\alpha}[\sqrt{-g}(1+\mathcal{L}_{m}
\delta g^{\alpha\nu}\right\} .
\end{array}
\end{equation}
Inserting eq. (30) in the variation (18) we get 
\begin{equation}
\begin{array}{c}
\delta\int d^{4}x\sqrt{-g}(R+R\mathcal{L}_{m}+\mathcal{L}_{m})=\int d^{4}x[\sqrt{-g}(1+\mathcal{L}_{m})R_{\mu\nu}-\frac{1}{2}g_{\mu\nu}(R+R\mathcal{L}_{m}+\mathcal{L}_{m})]\delta g^{\mu\nu}+\\
\begin{array}{c}
\end{array}\\
\int d^{4}x\left\{ g_{\mu\nu}\partial^{\alpha}\partial_{\alpha}[\sqrt{-g}(1+\mathcal{L}_{m})]-g_{\alpha\nu}\partial^{\mu}\partial_{\alpha}[\sqrt{-g}(1+\mathcal{L}_{m})]\delta g_{\mu\nu}\right\} +d^{4}x\left\{ (1+R)\delta(\sqrt{-g}\mathcal{L}_{m})\right\} .
\end{array}
\end{equation}
This variation is equal to zero for \cite{key-25,key-26}

\begin{equation}
R_{\mu\nu}-\frac{1}{2}Rg_{\mu\nu}=-\mathcal{L}_{m}R_{\mu\nu}+\left(\bigtriangledown_{\mu}\bigtriangledown_{\nu}-g_{\mu\nu}\square\right)\mathcal{L}_{m}+\frac{\left(1+R\right)}{2}T_{\mu\nu}^{(m)}
\end{equation}
which are the Einstein field equations modified by direct coupling
between the Ricci curvature scalar and the matter Lagrangian. In fact,
the standard stress-energy tensor \cite{key-27}

\begin{equation}
T_{\mu\nu}^{(m)}\equiv\frac{-2}{\sqrt{-g}}\frac{\delta(\sqrt{-g}\mathcal{L}}{\delta g^{\mu\nu}}
\end{equation}
has been introduced in the modified field equations (32). Writing
down explicitly the Einstein tensor and introducing a ''total''
stress-energy tensor \cite{key-25,key-26,key-28} 
\begin{equation}
T_{\mu\nu}^{(tot)}\equiv\frac{1}{(1+\mathcal{L}_{m})}\left[\left(\bigtriangledown_{\mu}\bigtriangledown_{\nu}-g_{\mu\nu}\square\right)\mathcal{L}_{m}+\frac{(1+R)}{2}T_{\mu\nu}^{(m)}-\frac{R\mathcal{L}_{m}}{2}g_{\mu\nu}\right]
\end{equation}
eqs. (32) can be put in the well known Einsteinian form 
\begin{equation}
G_{\mu\nu}=\frac{1}{2}T_{\mu\nu}^{(tot)},
\end{equation}
in which a \emph{curvature} contribution \cite{key-28} is added and
mixed to the \emph{material} one. In other words, the high order terms
contribute, like sources, to the modified field equations and have
to be considered like \emph{effective fields} (\cite{key-28} for
details). The condition of energy conservation \cite{key-25,key-26,key-27}

\begin{equation}
\bigtriangledown^{\mu}G_{\mu\nu}=0
\end{equation}
can be inserted in eqs. (35) and (34), obtaining 
\begin{equation}
\bigtriangledown^{\mu}T_{\mu\nu}^{(m)}=\frac{1}{R+1}(g_{\mu\nu}
\end{equation}
Now, we can introduce the well known stress-energy tensor of a perfect
fluid \cite{key-27} 
\begin{equation}
T_{\mu\nu}^{(m)}\equiv(\epsilon+p)u_{\mu}u_{\nu}-pg_{\mu\nu},
\end{equation}
in order to test the motion of test particles \cite{key-25,key-26},
where $\epsilon$ is the proper energy density, $p$ the pressure
and $u_{\mu}$ the fourth-velocity of the particles. This is the simplest
version of a stress-energy tensor for the matter, concerning inchoerent
matter, and it is considered a good approximation in astrophysics
frameworks \cite{key-25,key-26,key-27}. Introducing the projector
operator \cite{key-25,key-26} 
\begin{equation}
P_{\mu\alpha}\equiv g_{\mu\alpha}-u_{\mu}u_{\alpha},\label{eq: po}
\end{equation}
we can apply the contraction $g^{\alpha\beta}P_{\mu\beta}$ to equation
(37), obtaining \cite{key-25,key-26} 
\begin{equation}
\frac{d^{2}x_{\mu}}{ds^{2}}+\Gamma_{\alpha\beta}^{\mu}\frac{dx_{\alpha}}{ds}\frac{dx_{\beta}}{ds}=F^{\alpha}
\end{equation}
Thus, we find the existence of an extra force \cite{key-25,key-26}
\begin{equation}
F^{\alpha}\equiv(\epsilon+p)^{-1}P^{\alpha\nu}\left[(\frac{1}{R+1})\left(\mathcal{L}_{m}+p\right)\bigtriangledown_{\mu}R+\bigtriangledown_{\mu}p\right]
\end{equation}
showing that the motion of test particles is \textbackslash{}emph\{non-geodesic\}.
This extra force is orthogonal to the four-velocity of test masses
\cite{key-25,key-26}

\begin{equation}
F^{\alpha}\frac{dx_{\alpha}}{ds}=0.\label{eq: ortogonali}
\end{equation}
The Newtonian limit in three dimensions of equation (40) reads \cite{key-25,key-26}

\begin{equation}
\overrightarrow{a}_{tot}=\overrightarrow{a}_{n}+\overrightarrow{a}{}_{ng}.\label{eq: vect}
\end{equation}
The total acceleration $\overrightarrow{a}_{tot}$ is given by the
ordinary Newtonian acceleration $\overrightarrow{a}_{n}$ plus the
repulsive acceleration $\overrightarrow{a}_{ng}$ due to the extra
(non-geodesic) force \cite{key-25,key-26}. Eq. ($\ref{eq: vect}$)
and a bit of three-dimensional geometry permit to write the Newtonian
acceleration $\overrightarrow{a}_{n}$ as \cite{key-25,key-26}

\begin{equation}
\overrightarrow{a}_{n}=\frac{1}{2}(a_{tot}^{2}-a_{n}^{2}-a_{ng}^{2})\frac{\overrightarrow{a}_{tot}}{a_{ng}a_{tot}}
\end{equation}
Considering the limit in which $\overrightarrow{a}{}_{ng}$ dominates
(i.e. $a_{n}\ll a_{tot}$) one gets

\begin{equation}
a_{n}\simeq\frac{a_{tot}\overrightarrow{a}_{tot}}{2a_{ng}}\left(1-\frac{a_{ng}^{2}}{a_{tot}^{2}}\right).
\end{equation}
Thus, the extra aceleration is given by \cite{key-25,key-26}

\begin{equation}
a_{0}\equiv\left[\frac{1}{2a_{ng}}(1-\frac{a_{ng}^{2}}{a_{tot}^{2}})\right]^{-1},\label{eq: aE}
\end{equation}
and combining eq. ($\ref{eq: aE}$) with eqs. (10) and (11) one gets

\begin{equation}
\frac{m_{g}}{m_{i}}=\sqrt{\frac{g_{0}}{g}}=\sqrt{\frac{r^{2}}{\frac{M_{g}}{2a_{ng}}(1-\frac{a_{ng}^{2}}{a_{tot}^{2}})}}
\end{equation}
with 
\begin{equation}
g_{0}=\frac{r^{2}}{\frac{1}{2a_{ng}}(1-\frac{a_{ng}^{2}}{a_{tot}^{2}})}.\label{eq: g con zero}
\end{equation}
Thus, we have shown that in our model the ratio between gravitational
and inertial mass is explained in an elegant, geometric way through
a direct coupling between the Ricci curvature scalar and the matter
Lagrangian which generates a non geodesic motion of test particles.

\section{Conclusions}

The assumption of an $R$-dependent inertial mass is in accordance
with the spirit of Mach principle and Einstein himself tried to implement
this hypothesis in the context of General Relativity \cite{key-29}.
The possible relation $m_{g}/m_{i}$ is deduced by comparing it with
Milgrom's rotational equation that is in perfect agreement with the
experimental data. However, the interpretation given here is different,
and it leaves unchanged in any point the second law of dynamics. Finally
we have shown that the ratio between gravitational and inertial mass
is explained in geometric way through a direct coupling between the
Ricci curvature scalar and the matter Lagrangian which generates a
non geodesic motion of test particles. Such a non geodesic motion
is due to the presence of the centrifugal extra force in eq. (42).

\section{Acknowledgements}

The authors thank an unknown referees for useful comments.


\begin{thebibliography}{10}
\bibitem{key-1}Mach E., 1883, Die Mechanik in ihrer Entwicklung Historisch-Kritisch
Dargerstellt, Leipzig: Brockhaus.

\bibitem{key-2}H. Ohanian, R. Ruffini \textquotedbl{}Gravitation
and Spacetime\textquotedbl{}, W.W. Norton, New York (1994).

\bibitem{key-3}J. V. Narlikar, Action at a distance and cosmology:
A historical perspective, Annual Review of Astronomy and Astrophysics,
41, (2003): 169-189.

\bibitem{key-4}J. V. Narlikar, Inertia and cosmology in Einstein's
relativity, in Relativity, Quanta and Cosmology, Einstein Centenary
Vol. II. Eds. M. Pantaleo and F. de Finis, Johnson Reprint Corporation,
493 (1979).

\bibitem{key-5}D. Sciama, On the Origin of Inertia, Monthly Notices
of the Royal Astronomical Society, Vol. 113 (1953):34- 42.

\bibitem{key-6}J. Barbour, H. Pfister, Mach's Principle: From Newton's
Bucket to Quantum Gravity, Birkhauser, 1995.

\bibitem{key-7}H. C. Rosu, Classical and Quantum Inertia: A Matter
of Principle, Gravitation and Cosmology vol. 5 No. 2 (1999) 81-91.

\bibitem{key-8}A. B. Arbuzov, L. A. Glinka, V. N. Pervushin, Higgs
Particle Mass in Cosmology, arXiv:0705.4672 {[}hep-ph{]}, 2008.

\bibitem{key-9}Y. N. Srivastava, J. Swain, A. Widom, An Argument
for Nonminimal Higgs Coupling to Gravity, arXiv:1110.5549 {[}gr-qc{]},
2011.

\bibitem{key-10}Y. N. Srivastava, A. Widom, Gravitational Decay Modes
of the Standard Model Higgs Particle, arXiv:hep-ph/0003311, 2000.

\bibitem{key-11}Einstein, A., The Meaning of Relativity. Four lectures
delivered at Princeton University, May, 1921, Princeton Univ. Press,
2004.

\bibitem{key-12}Brans C.H. \textquotedbl{}Mach's Principle and the
Locally Measured Gravitational Constant in General Relativity\textquotedbl{}
Phys. Rev. Vol. 125, 388 (1962).

\bibitem{key-13}F. Darabi \textquotedbl{}Is flat rotation curve a
sign of cosmic expansion?\textquotedbl{} MNRAS 433, (2013).

\bibitem{key-14}F. Zwicky, Spectral displacement of extra galactic
nebulae, Helv. Phys. Acta. 6, 110-127 (1933).

\bibitem{key-15}M. Milgrom \textquotedbl{}A modification of the Newtonian
dynamics as a possible alternative to the hidden mass hypothesis\textquotedbl{}.
Astrophysical Journal 270: 365-370 (1983).

\bibitem{key-16}M. Milgrom \textquotedbl{}A modification of the Newtonian
dynamics - Implications for galaxies\textquotedbl{} Astrophysical
Journal 270: 371-389 (1983).

\bibitem{key-17}M. Milgrom, ``MOND theory'', Canadian Journal of
Physics, e-First Article: pp. 1-12, doi: 10.1139/cjp-2014-0211 (2014).

\bibitem{key-18}M. Milgrom \textquotedbl{}MD or DM? Modified dynamics
at low accelerations vs dark matter\textquotedbl{}. Proceedings of
Science (2011).

\bibitem{key-19}V.A. De Lorenci, M. Faundez-Abans, J.P. Pereira \textquotedbl{}Testing
the Newton second law in the regime of small accelerations\textquotedbl{}
Astronomy \& Astrophysics, Vol. 503, N. 1 (2009).

\bibitem{key-20}K.G. Begeman, A.H. Broeils, R.H. Sanders \textquotedbl{}Extended
rotation curves of spiral galaxies: dark haloes and modified dynamics\textquotedbl{}
MNRAS 249, 523 (1991).

\bibitem{key-21}J. Bekenstein, M. Milgrom \textquotedbl{}Does the
missing mass problem signal the breakdown of Newtonian gravity?\textquotedbl{}
Astrophysical Journal 286, 7-14 (1984).

\bibitem{key-22}W. J. G de Blok, S. S. McGaugh \textquotedbl{}Testing
Modified Newtonian Dynamics with Low Surface Brightness Galaxies:
Rotation Curve Fits\textquotedbl{} Astrophysical Journal 508, 132-140
(1998).

\bibitem{key-23}Mortlock, D. J., and Turner, E. L.\textquotedbl{}Gravitational
lensing in modified Newtonian dynamics\textquotedbl{} MNRAS 327, 557
(2001).

\bibitem{key-24}Das S., Patitsas S. N. \textquotedbl{}Can MOND type
hypotheses be tested in a free fall laboratory environment?\textquotedbl{}
Physical Review D. 87, 107101 (2013).

\bibitem{key-25}O. Bertolami, C. G. Bohmer, T. Harko and F. S. M.
Lobo, Phys. Rev. D 75, 104016 (2007).

\bibitem{key-26}C. Corda, Int. J. Theor. Phys. 47, 2679 ((2008).

\bibitem{key-27}L. Landau L and E. Lifsits, \emph{Classical Theory
of Fields} (3rd ed.). London: Pergamon. ISBN 0-08-016019-0. Vol. 2
of the Course of Theoretical Physics (1971).

\bibitem[28]{key-28}C. Corda, N. Adv. Phys. 7, 1, 67 (2013). 

\bibitem[29]{key-29}A. Pais, \textit{Subtle Is the Lord: The Science
and the Life of Albert Einstein''}, Oxford University Press (2005).\end{thebibliography}
\end{document}